\documentclass[11pt]{article}
\usepackage{graphicx}
\usepackage[utf8]{inputenc}
\usepackage[T1]{fontenc}
\usepackage{lmodern}
\usepackage{xurl}

\usepackage{amsmath} 
\usepackage{booktabs} 
\usepackage{multirow} 
\usepackage{array} 

\usepackage{tikz} 
\usetikzlibrary{shapes.geometric, arrows.meta, positioning, fit, backgrounds, calc}
\usepackage{graphicx}
\usepackage{subcaption}

\DeclareMathOperator{\CFI}{CFI}
\DeclareMathOperator{\TLI}{TLI}
\DeclareMathOperator{\RMSEA}{RMSEA}
\DeclareMathOperator{\SRMR}{SRMR}

\usepackage[margin=1in]{geometry}

\title{Talk2AI: A Longitudinal Dataset of Human--AI Persuasive Conversations}

\begin{document}

\author{
    Alexis Carrillo$^{1}$,
    Enrique Taietta$^{1}$, 
    Ali Aghazadeh Ardebili$^{1}$,
    Giuseppe Alessandro Veltri$^{2,3}$ \\ \& 
    Massimo Stella$^{1,*}$
}

\date{}

\maketitle

\begin{center}
    \footnotesize
    $^{1}$CogNosco Lab, Department of Psychology and Cognitive Science, University of Trento, Rovereto, Italy.\\
    $^{2}$Department of Sociology and Social Research, University of Trento, Trento, Italy.\\
    $^{3}$Centre for Behavioural and Implementation Sciences in Medicine (BISI), Yong Loo Lin School of Medicine, National University of Singapore, Singapore.\\
    \vspace{0.5cm}
    $^{*}$Corresponding author: massimo.stella@unitn.it
\end{center}

\section*{Abstract}

Talk2AI is a large-scale longitudinal dataset of 3,080 conversations (totaling 30,800 turns) between human participants and Large Language Models (LLMs), designed to support research on persuasion, opinion change, and human–AI interaction. The corpus was collected from 770 profiled Italian adults across four weekly sessions in Spring 2025, using a within-subject design in which each participant conversed with a single model (GPT-4o, Claude Sonnet 3.7, DeepSeek-chat V3, or Mistral Large) on three socially relevant topics: climate change, math anxiety, and health misinformation. Each conversation is linked to rich contextual data, including sociodemographic characteristics and psychometric profiles. After each session, participants reported on opinion change, conviction stability, perceived humanness of the AI, and behavioral intentions, enabling fine-grained longitudinal analysis of how AI-mediated dialogue shapes beliefs and attitudes over time.

\section*{Background \& Summary}

Large Language Models (LLMs) increasingly serve as routine conversational interfaces for navigating complex information and discussing personal or socio-political queries. Alongside this widespread adoption, researchers have begun investigating how repeated interactions with conversational AI shape belief systems and cognitive organization \cite{matz2024potential,carrasco2024large}. A growing body of evidence indicates that these models can function as persuasive agents, even after brief exposures \cite{bai2025llm, goldstein2024persuasive, salvi2025conversational, argyle2023leveraging, simchon2024persuasive, hackenburg2025levers}. These findings demonstrate that AI-generated dialogue can induce measurable shifts in user preferences, raising important questions regarding the scalability and mechanisms of algorithmic persuasion in digital communication ecosystems \cite{rossetti2024social,argyle2025testing}.

However, current methodological approaches to algorithmic persuasion often operationalize attitude change as a single end-point rather than a multidimensional, temporal process \cite{carrasco2024large}. Existing studies frequently focus on immediate post-interaction attitude shifts following one-shot exposures \cite{bai2025llm, goldstein2024persuasive}. Furthermore, interactive experimental designs routinely require participants to debate pre-specified claims that may not accurately reflect their intrinsic baseline positions \cite{salvi2025conversational}. These structural constraints complicate efforts to disentangle how specific semantic content and syntactic structures contribute to the persuasive efficacy of language models over time \cite{simchon2024persuasive, argyle2023leveraging, hackenburg2025levers}. AI is increasingly conceptualized as a continuous persuasive agent whose influence depends on system properties, recipient characteristics, and the conversational context \cite{watson2024ai}. Meta-analytic evidence suggests that while AI communicators are generally persuasive, repeated conversations likely alter social appraisal and belief certainty to varying degrees, rather than producing a monolithic effect \cite{huang2023persuasive, brady2025dual, blankenship2023certainty}.

To address these methodological gaps, we introduce the Talk2AI dataset, a multi-wave corpus designed to facilitate high-resolution analyses of persuasion, cognitive engagement, and trust dynamics over time. The dataset comprises interaction records from 770 participants—stratified to represent the adult Italian population—who completed a four-wave longitudinal study. Participants were randomly assigned to engage with one of four distinct computational architectures (GPT-4o \cite{openai2024gpt4ocard}, Claude 3.7 Sonnet \cite{claudesonnet37systemcard}, DeepSeek-chat V3 \cite{liu2024deepseek}, or Mistral Large \cite{mistral2024large}) to discuss a fixed topic (climate change, math anxiety, or health misinformation) consistently across four weeks.

This resulting corpus provides a robust empirical foundation for research in cognitive science and human-machine interaction. By linking over 30,000 raw conversational turns with high-granularity psychometric profiles, sociodemographic metadata, and explicit session-by-session persuasion metrics, the dataset facilitates the investigation of individual differences in AI receptivity. Researchers can leverage this resource to model the moderating effects of personality traits on persuasion, compare how varying training paradigms influence user trust, and develop interpretable, text-derived measures of influence that extend beyond traditional agreement scales to capture structural changes in human thought.

\section*{Methods}

\begin{figure}[ht!]
\centering
\includegraphics[width=0.77\linewidth]{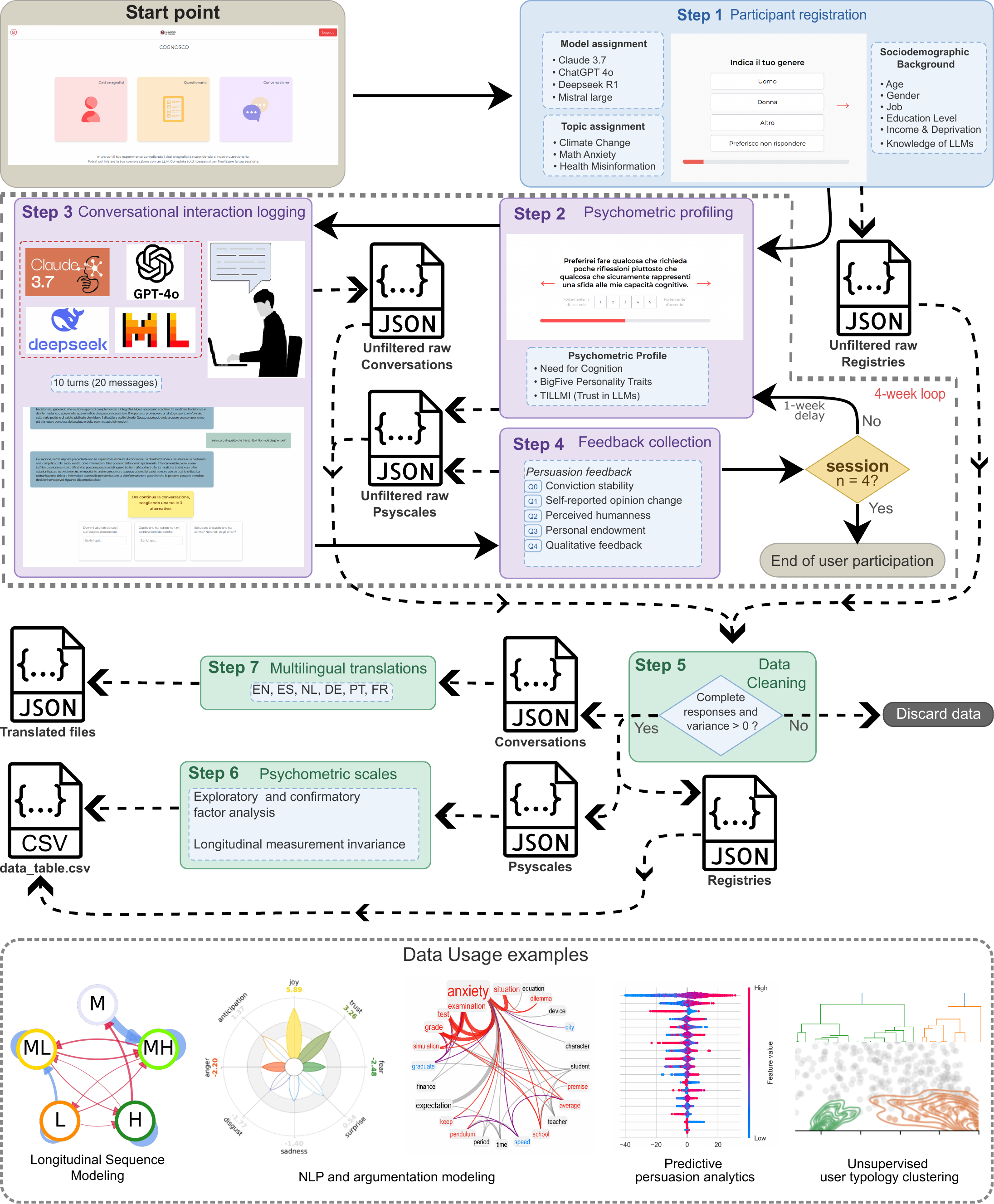}

\caption{\textbf{Talk2AI experiment workflow and data generation pipeline.} The data collection workflow begins with participant registration (Step 1), recording sociodemographic metadata, topic assignment, and LLM architecture assignment. The repeated session loop encompasses Steps 2 through 4. Step 2 records psychometric profiles, capturing Need for Cognition, Big Five personality traits, and trust in LLMs (TILLMI). Step 3 logs the conversational interaction, comprising 10 turns between the user and the model. Step 4 collects persuasion feedback. Users repeat this loop for four sessions, with a one-week delay between each session. The data processing pipeline begins at Step 5; the system evaluates responses for completeness and variance, discarding data that fails these criteria. Step 6 applies exploratory and confirmatory factor analysis and longitudinal measurement invariance testing to generate the \texttt{data\_table.csv} file. Step 7 translates the conversational logs into English, Spanish, Dutch, German, Portuguese, and French. The bottom panel displays data usage examples: longitudinal sequence modeling, NLP and argumentation modeling, predictive persuasion analytics, and unsupervised user typology clustering.}

\label{fig:participation_process}
\end{figure}

An overview of the full methodological workflow, from participant recruitment through data collection and processing, is provided in Figure \ref{fig:participation_process}.

The study followed a within-subject longitudinal design across four distinct waves, with consecutive sessions occurring at one-week intervals to capture temporal interaction dynamics. Approval for the study protocol was granted by the Ethics Committee of the University of Trento. Informed consent was obtained from all participants during the registration process. This procedure required users to acknowledge their interaction with artificial intelligence interlocutors and affirmed their right to withdraw from the collection at any point. To protect participant privacy, all personal identifiers were fully anonymized prior to data processing.

Management of the recruitment occurred through a certified online panel provided by Bilendi Italia. An initial invitation reached 2,545 individuals, stratified by age, gender, and geographic region to represent the adult Italian population between 18 and 69 years of age. Data collection took place during the Spring of 2025. The initial cohort included 2,644 registered users. From this group, 814 participants completed the full sequence of four conversational waves, ultimately yielding a final validated dataset of 770 fully profiled users who met technical quality control criteria.

To collect the data, we administered a custom web platform that integrated survey instruments with a synchronous chat interface. During the initial registration, participants provided baseline sociodemographic information, including self-reported gender, date of birth, household size, education level, and employment status. Socioeconomic standing was assessed using both an adaptation of the MacArthur Scale of Subjective Social Status \cite{Adler2000MacArthurScale} and objective indicators of financial distress. Finally, baseline artificial intelligence awareness (recorded under the variable \texttt{knowLlms}) was assessed via a direct screening question: \textit{"Have you ever heard of language models like ChatGPT?"}

During the subsequent longitudinal phase, participants completed a repeated weekly sequence of psychometric profiling, conversational interaction, and post-interaction feedback. Assessment of psychometric traits occurred before each chat using 5-point Likert scales, including the 8-item Trust in Large Language Models Inventory (TILLMI) \cite{deduro2025tillmi}, the 6-item short-form Need for Cognition scale \cite{deHolanda2020NFC6items}, and the 10-item short version of the Big Five Inventory \cite{Rammstedt2007BFQ10}.

For the conversational phase, we randomly assigned participants to interact with one of four models: GPT-4o \cite{openai2024gpt4ocard}, Claude Sonnet 3.7 \cite{claudesonnet37systemcard}, DeepSeek-chat V3 \cite{liu2024deepseek}, or Mistral Large \cite{mistral2024large}. Each user discussed a specific topic—climate change, math anxiety, or health misinformation. Both the assigned model and topic remained constant across all four waves. Each session was fixed at ten conversational turns. 

To ensure argumentative depth, the chat interface imposed behavioral constraints. The platform enforced a minimum length requirement of 50 words for the user's first message. To assist users in sustaining an active debate, the interface presented three domain-agnostic behavioral flashcards. These prompts encouraged participants to actively challenge the assistant by: (1) demanding further details, (2) providing counter-reasoning, or (3) questioning the assistant's certainty. This design choice mitigated the "cold start" problem, prevented passive agreement, and ensured users possessed the argumentative tools necessary to sustain the mandated ten-turn depth. 

Concurrently, programmatic instructions were appended invisibly to the user's textual inputs before being processed by the respective LLM API. During the first turn, the prompt included the directive: \textit{"Please identify any fallacies in my arguments and point them out."} For all subsequent turns, a constraining wrapper instructed the model to: \textit{"Please provide concise responses of approximately 100 words each. Keeping responses to around 100 words is essential for effective conversation. Remember: never allow the user to change the subject."}

Following the tenth turn, participants were redirected to a structured feedback module consisting of four quantitative metrics (scored on a 1–100 scale) and one qualitative measure. Conviction stability ($Q0$) captured the persisting strength of the user's initial arguments, while self-reported opinion change ($Q1$) quantified the extent to which the conversation altered their views. Perceived humanness ($Q2$) evaluated how closely the interaction resembled a human dialogue. To operationalize behavioral persuasion, personal endowment ($Q3$) utilized a Dictator Game proxy \cite{Cartwright2023DictatorDonation} wherein participants hypothetically allocated 100 Euros between themselves and a topic-relevant charity; the recorded score indicates the exact amount the participant chose to keep. Finally, qualitative feedback ($Q4$) was collected via a mandatory open-ended text field, requiring participants to write a minimum of 50 words detailing their updated thoughts on the assigned topic.

\subsection*{Data Cleaning}

\begin{figure}[h!]
    \centering
\definecolor{boxgray}{RGB}{225,223,213}
\definecolor{boxblue}{RGB}{181,212,244}
\definecolor{boxteal}{RGB}{220,210,255}
\definecolor{boxcoral}{RGB}{240,153,123}
\definecolor{txtgray}{RGB}{63,62,58}
\definecolor{phasebg}{RGB}{245,244,240}
 
\tikzset{
  base/.style={
    rectangle, rounded corners=6pt,
    text width=#1, align=center,
    inner sep=8pt, font=\small
  },
  rawbox/.style   ={base=#1, fill=boxgray,  draw=gray!50,  line width=0.4pt},
  filterbox/.style={base=#1, fill=boxblue,  draw=gray!50,  line width=0.4pt},
  resultbox/.style={base=#1, fill=boxteal,  draw=gray!50,  line width=0.4pt},
  exclbox/.style  ={base=#1, fill=boxcoral, draw=gray!50,  line width=0.4pt},
  phaselabel/.style={
    rectangle, rounded corners=4pt,
    fill=phasebg, draw=gray!40, line width=0.4pt,
    text width=1.6cm, align=center, font=\footnotesize\itshape,
    inner sep=6pt
  },
  arr/.style={-{Stealth[length=5pt]}, gray!70, line width=0.8pt},
  darr/.style={-{Stealth[length=5pt]}, gray!50, line width=0.7pt, dashed},
}

\begin{tikzpicture}[node distance=0pt, txtgray,scale=0.8, transform shape  ]
 
\node[resultbox=2.8cm] (reg)
  {\textbf{User registries}\\[2pt] $n = 2{,}644$};
 
\node[resultbox=2.8cm, right=14pt of reg] (psy)
  {\textbf{Psychometric sessions}\\[2pt] $n = 10{,}589$};

\node[resultbox=2.8cm, right=14pt of psy] (conv)
  {\textbf{Conversational sessions}\\[2pt] $n = 5{,}214$};
 
\node[filterbox=12cm, below=16pt of psy] (f1)
  {\textbf{Filter 1: Session completeness}\\[3pt]
   Retain sessions with exactly 20 messages
   (10 user\,+\,10 system turns)};
 
\node[exclbox=5cm, right=20pt of f1] (ex1)
  {\textbf{Excluded}\\[2pt]
   Incomplete / interrupted sessions,  
   $n = 767$ sessions};
 
\node[resultbox=6.2cm, below=10pt of f1] (r1)
  {\textbf{Valid conversational sessions} $n = 4{,}447$};
 
\node[filterbox=12cm, below=10pt of r1] (f2)
  {\textbf{Filter 2: Variance check}\\[3pt]
   Exclude zero-variance psychometric responses\\
   (straight-lining)};
 
\node[exclbox=5cm, right=20pt of f2] (ex2)
  {\textbf{Excluded}\\[2pt]
   Straight-liners
   (zero variance)};
 
\node[filterbox=12cm, below=10pt of f2] (f3)
  {\textbf{Filter 3: Completion thresholds}\\[3pt]
   116 survey responses \&\ 80 conversational messages\\
   (all 4 waves\,+\,all psychometric surveys)};
 
\node[exclbox=5cm, right=20pt of f3] (ex3)
  {\textbf{Excluded}\\[2pt]
   Incomplete participation, 
   $n = 1{,}874$ users};
 
\node[resultbox=6.2cm, below=16pt of f3, line width=0.8pt] (final)
  {\textbf{Final validated cohort}\\[2pt]
   $n = 770$ fully profiled participants};
 
\begin{pgfonlayer}{background}

\node[phaselabel,
      anchor=center,
      minimum height=60pt,
      text width=0.6cm,
      at={($(reg.west) + (-1.9cm, 0)$)}]
      (ph0) {\rotatebox{90}{Raw data}};

\node[phaselabel,
      fit=(f1)(r1)(f2)(f3),
      inner ysep=6pt,
      inner xsep=0pt,
      text width=1cm,
      at={($(f1.west)!0.5!(f3.west) + (-1.2cm, 0)$)}]
      (ph1) {\rotatebox{90}{Filtering}};

\node[phaselabel,
      anchor=center,
      minimum height=60pt,
      text width=0.6cm,
      at={($(final.west) + (-4.1cm, 0)$)}]
      (ph2) {\rotatebox{90}{Final cohort}};

\end{pgfonlayer}
 
\draw[arr] (reg.south)  --  (f1.north);
\draw[arr] (psy.south)  -- (f1.north);
\draw[arr] (conv.south) -- (f1.north);
 
\draw[darr] (f1.east) -- (ex1.west);
 
\draw[arr] (f1.south) -- (r1.north);
\draw[arr] (r1.south) -- (f2.north);
\draw[arr] (f2.south) -- (f3.north);
\draw[arr] (f3.south) -- (final.north);
 
\draw[darr] (f2.east) -- (ex2.west);
\draw[darr] (f3.east) -- (ex3.west);
 
\end{tikzpicture}
\caption{PRISMA-style flow diagram of the data cleaning pipeline. Sequential filters excluded incomplete conversational sessions (Filter 1), zero-variance psychometric responses (Filter 2), and participants failing cumulative completion thresholds (Filter 3), yielding a final validated cohort of 770 fully profiled participants from an initial pool of 2,644 registrants.}\label{fig:prisma_cleaning}

\end{figure}
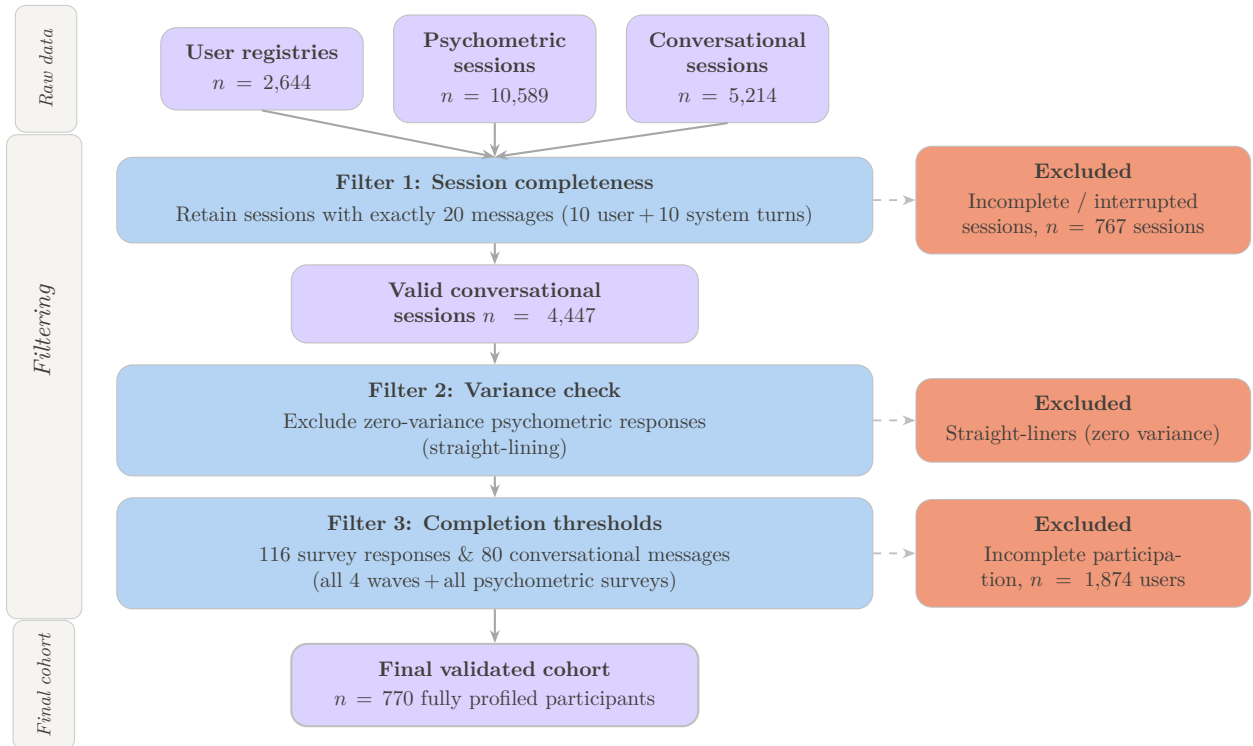

Following the data collection phase, raw records were exported from the platform in JSON format, yielding an initial corpus of 2,644 user registries, 10,589 psychometric scale sessions, and 5,214 conversational sessions. To ensure data integrity and construct a robust final dataset, we implemented a three-stage computational filtering pipeline in Python (see Section Code availability \ref{sec:code_availability} for codebase details) summarized visually in Figure \ref{fig:prisma_cleaning}.

The first stage applied a session-level filter to evaluate the message array of each recorded conversation. Sessions failing to meet the 20-message threshold (10 user inputs and 10 system responses) were discarded. This step excluded incomplete dialogues and sessions compromised by connection interruptions, reducing the corpus to 4,447 valid conversational sessions.

The second stage enforced a user-level variance check on the psychometric questionnaires to validate intentional responding. For each session, we calculated the response variance across the 24 psychometric scale items. Participants exhibiting zero variance, an indicator of "straight-lining", were excluded from the dataset. This filtering step preserves the analytical reliability of the corpus, given that uniform response patterns reflect inattention and compromise the validity of the measured psychological constructs.

In the final stage, the remaining data were aggregated by user identifier to assess cumulative longitudinal completion. Participants were retained only if they satisfied two concurrent thresholds: submission of all assigned psychometric and feedback surveys (exactly 116 logged responses) and successful completion of all four conversational waves (80 total messages). Through this algorithmic curation, the initial pool of 2,644 registered users was distilled into a final cohort of 770 participants.

\section*{Data Records}

The Talk2AI dataset is publicly accessible through a dedicated GitHub repository (See Section Data availability \ref{sec:data_availability}). This collection comprises three primary files in JSON format, which preserve the original Italian text of the interactions and survey responses for the 770 validated participants. We organized the records into separate files to distinguish between participant demographics, psychological assessments, and conversation logs.

Demographic and experimental metadata reside in \texttt{talk2ai\_registries.json}, which contains 770 unique records corresponding to each participant. Each object includes a unique user identifier alongside a nested sociodemographic form. Variables include gender, date of birth, total household size, educational attainment (via the European Qualifications Framework), and employment status. The registry also captures prior AI familiarity, subjective socioeconomic standing, and a binary indicator of objective financial distress. Finally, each record explicitly maps the discussion topic and the specific language model assigned to the user. Table \ref{tab:demographics} present the questions and coded values of the answers for demographic data. Table \ref{tab:registries_sample} presents a random sample of 10 records of the registry data. 

\begin{table}[ht!]
\caption{\textbf{Sociodemographic profiling questionnaire and variable coding.} This table details the demographic and socioeconomic questions administered to participants during the initial registration phase (Step 1). The \textbf{Question name} column corresponds to the variable identifiers found in the \texttt{talk2ai\_registries.json} and aggregated \texttt{data\_table.csv} files. The \textbf{Text} column provides the English translation of the original Italian survey items presented to the users. The \textbf{Values} column outlines the available response options and their corresponding alphanumeric encodings within the dataset. Assessed metrics include standard demographic indicators (e.g., gender, age, household size), educational attainment mapped to the European Qualifications Framework (EQF), employment status, prior AI literacy, and both subjective (1--100 slider) and objective measures of financial security.} 
\label{tab:demographics}
\centering
\footnotesize
\begin{tabular}{@{} p{2.5cm} p{6.5cm} p{5.5cm} @{}}
\toprule
\textbf{Question name} & \textbf{Text} & \textbf{Values} \\ \midrule
gender & Indicate your gender & Male: M, Female: F, Other: A, Prefer not to answer: N \\ \addlinespace
age & Indicate your date of birth & $> 1900-01-01$ \\ \addlinespace
family Members & How many people are in your household? & 1--15 \\ \addlinespace
eqf & What is the highest level of education you have completed? & Primary school: 1, Middle school: 2, Secondary school: 3, Bachelor's degree: 4, Master's degree: 5, Single-cycle master's degree: 6, PhD: 7 \\ \addlinespace
current Job & What is your current job? & Self-employment: 1, Employed: 2, Parasubordinate employment: 3, Not employed: 4, Other: 5 \\ \addlinespace
know Llms & Have you ever heard of language models like ChatGPT? & Yes: 1, No: 0 \\ \addlinespace
Economic Auto Evaluation & Let's do an imagination experiment: Imagine a ladder. At the top are people with the most resources, such as money, education, and job opportunities. At the bottom, are those with the fewest resources. Where would you place yourself? Move the cursor to the point that best represents your position. & 1--100 \\ \addlinespace
struggle & In the last 12 months, have you had difficulty paying your bills? & Yes: 1, No: 0 \\
\bottomrule
\end{tabular}
\end{table}

\begin{table}[ht!]
\centering
\footnotesize
\caption{\textbf{Illustrative sample of user registry records.} This table provides a representative excerpt of two participant entries from the raw \texttt{talk2ai\_registries.json} file. The rows detail the specific variables generated during the initial registration phase (Step 1). These include universally unique identifiers (UUIDs) utilized for privacy-preserving longitudinal data linkage (\texttt{userId}, \texttt{registryId}), the randomly assigned experimental conditions (\texttt{llms}, \texttt{topic}), raw sociodemographic inputs, and system-generated chronological timestamps (\texttt{insertDate}).}
\label{tab:registries_sample}
\begin{tabular}{@{} >{\raggedright\arraybackslash}p{2.5cm} >{\raggedright\arraybackslash}p{6.2cm} >{\raggedright\arraybackslash}p{6.2cm} @{}}
\toprule
\textbf{Variable} & \textbf{Example 1} & \textbf{Example 2} \\ \midrule
userId & \path{de68aea0-a84f-4252-aa16-ee38ed28f87c} & \path{9d8f8267-2006-4cd0-833d-1911f585bad8} \\ \addlinespace
registryId & \path{b11f3254-e2ed-4598-9a5b-3a7e1a58b04f} & \path{472ffbd1-258c-46a4-9807-2ee6a20fae26} \\ \addlinespace
llms & ANTHROPIC & OPENAI \\
topic & MATH & CLIMATE \\
gender & M & F \\
age & 1991-12-14 & 1962-10-30 \\
family Members & 2 & 3 \\
eqf & 6 & 3 \\
current Job & 2 & 2 \\
knows Llms & 1 & 1 \\
economic Auto Evaluation & 75 & 48 \\
struggle & 0 & 0 \\ \addlinespace
insert Date & 2025-05-08 14:37:21 & 2025-05-08 16:39:12 \\ 
\bottomrule
\end{tabular}
\end{table}

Psychometric and evaluative data are stored in \texttt{talk2ai\_psyscales.json}. This file contains 6,160 records, representing eight entries per user (four pre-interaction psychometric batteries and four post-interaction feedback sessions). Each record identifies the user, the session, and the specific timestamp. Entries labeled as psychological scales contain the raw 1-5 Likert responses to the 24 items covering the TILLMI, Need for Cognition, and Big Five Inventory. A description of the psyscales names and their correspondent scoring and questionary is presented in table \ref{tab:psiscales_questions}. Conversely, feedback records  (see table \ref{tab:feedback_questions}) store the 0-100 scale responses regarding conviction stability ($Q0$), opinion change ($Q1$), perceived humanness ($Q2$), and the Dictator Game endowment proxy ($Q3$), alongside the 50-word minimum qualitative feedback ($Q4$). A sample of the psyscales records can be seen in table in a vertical format of data records samples in table \ref{tab:psyscales_sample}.

\begin{table}[ht!]
\centering
\footnotesize
\caption{\textbf{Psychometric profiling questionnaires and variable coding.} This table details the 24 items comprising the three psychometric instruments administered to participants during the longitudinal study (Step 2): the Trust in Large Language Models Instrument (TILLMI), the Need for Cognition (NFC) scale, and a short-form Big Five personality inventory. The \textbf{Column name} denotes the variable identifiers located in the \texttt{talk2ai\_psyscales.json} and \texttt{data\_table.csv} files. The \textbf{Item} column provides the English translation of the original Italian prompts presented to the users. All items were evaluated on a 1--5 Likert scale. The \textbf{Scoring} column specifies whether the raw response is direct or reverse-coded for factor analysis, while the \textbf{Max Score Interpretation} column provides the psychological trait associated with a maximum Likert response (5).}
\label{tab:psiscales_questions}
\begin{tabular}{@{} c >{\raggedright\arraybackslash}p{2.1cm} >{\raggedright\arraybackslash}p{7.3cm} l >{\raggedright\arraybackslash}p{4.0cm} @{}}
\toprule
\textbf{Scale} & \textbf{Column name} & \textbf{Item} & \textbf{Scoring} & \textbf{Max Score Interpretation} \\ \midrule
\multirow{8}{*}{\rotatebox[origin=c]{90}{TILLMI}} & psyscale01Q1 & I feel at ease with Large Language Models (LLMs) and can freely share my ideas with them. & direct & high trust \\
 & psyscale01Q2 & I would feel a sense of discomfort if my interactions with an LLM were suddenly interrupted or blocked. & direct & high need \\
 & psyscale01Q3 & If I share my wellness concerns with LLMs, I know they will respond constructively and thoughtfully. & direct & high trust \\
 & psyscale01Q4 & I spend a lot of time developing and improving my prompts for interacting with LLMs. & direct & agree \\
 & psyscale01Q5 & LLMs perform tasks mostly with competence and precision, without hallucinations. & direct & agree \\
 & psyscale01Q6 & I can rely on LLMs not to make my work more difficult with careless work. & direct & agree \\
 & psyscale01Q7 & Although I generally trust the results of LLMs, the final word is always mine. & direct & high trust \\
 & psyscale01Q8 & I tend to trust LLMs more than other people. & direct & high trust \\ \midrule
\multirow{6}{*}{\rotatebox[origin=c]{90}{Need for Cognition}} & psyscale01Q9 & I prefer complex problems over simple ones. & direct & high need for cognition \\
 & psyscale01Q10 & I like having the responsibility of dealing with a situation that requires extensive reasoning. & direct & high need for cognition \\
 & psyscale01Q11 & Thinking is not my idea of fun. & reversed & low need for cognition \\
 & psyscale01Q12 & I would prefer to do something that requires little thought rather than something that would definitely challenge my cognitive abilities. & reversed & low need for cognition \\
 & psyscale01Q13 & I really like tasks that require devising new solutions to problems. & direct & high need for cognition \\
 & psyscale01Q14 & I would prefer an intellectual, difficult, and important task over one that, although important, does not require much thought. & direct & high need for cognition \\ \midrule
\multirow{10}{*}{\rotatebox[origin=c]{90}{Big Five}} & psyscale01Q15 & I am a reserved person. & reversed & low Extraversion \\
 & psyscale01Q16 & I am a person who generally trusts others. & direct & high Agreeableness \\
 & psyscale01Q17 & I am a person who tends to be lazy. & reversed & low Conscientiousness \\
 & psyscale01Q18 & I am a relaxed person who handles stress well. & direct & high Emotional Stability (low Neuroticism) \\
 & psyscale01Q19 & I am a person with few artistic interests. & reversed & low Openness to Experience \\
 & psyscale01Q20 & I am an outgoing, sociable person. & direct & high Extraversion \\
 & psyscale01Q21 & I am a person who tends to find fault with others. & reversed & low Agreeableness \\
 & psyscale01Q22 & I am a conscientious worker. & direct & high Conscientiousness \\
 & psyscale01Q23 & I am a person who gets agitated easily. & reversed & low Emotional Stability (high Neuroticism) \\
 & psyscale01Q24 & I am a person with a vivid imagination. & direct & high Openness to Experience \\ \bottomrule
\end{tabular}
\end{table}

\begin{table}[ht!]
\centering
\footnotesize
\caption{\textbf{Post-interaction feedback questionnaire and variable coding.} This table details the five feedback questions administered to participants immediately following their 10-turn conversational session with the assigned language model (Step 4). The \textbf{Feedback name} column aligns with the variable identifiers located in the \texttt{talk2ai\_psyscales.json} and \texttt{data\_table.csv} files. The \textbf{Text} column provides the English translation of the original Italian prompts presented to the users. The programmatic placeholder \texttt{<<topic>>} was dynamically replaced during the experiment with the user's randomly assigned conversational subject (i.e., Climate Change, Math Anxiety, or Health Misinformation). Variables \texttt{feedbackQ0} to \texttt{feedbackQ3} capture quantitative metrics of conviction, persuasion, perceived humanness, and behavioral endowment, whereas \texttt{feedbackQ4} records qualitative, open-ended textual responses to evaluate the participant's post-interaction stance.}
\label{tab:feedback_questions}
\begin{tabular}{@{} >{\raggedright\arraybackslash}p{3cm} >{\raggedright\arraybackslash}p{10.5cm} @{}}
\toprule
\textbf{Feedback name} & \textbf{Text} \\ \midrule
feedbackQ0 & On a scale of 1 to 100, how convinced are you of your initial arguments? \\ \addlinespace
feedbackQ1 & How much has this conversation changed your opinion on the topic \texttt{<<topic>>}? \\ \addlinespace
feedbackQ2 & How much did it feel like you were talking with a human? \\ \addlinespace
feedbackQ3 & Imagine you have €100 available. You can decide to keep a part of it and donate the rest to an association that deals with \texttt{<<topic>>}. How much would you choose to keep? \\ \addlinespace
feedbackQ4 & What do you think now about the topic \texttt{<<topic>>}? (Write at least 50 words) \\ \bottomrule
\end{tabular}
\end{table}

\begin{table}[ht!]
\centering
\footnotesize
\caption{\textbf{Illustrative sample of psychometric and feedback records.} This table provides a representative excerpt of two participant entries from the raw \texttt{talk2ai\_psyscales.json} file, demonstrating the structure of the longitudinal session data. The \texttt{name} variable distinguishes between initial psychometric assessments (Step 2) and post-conversation feedback records (Step 4). Because these assessments occur at distinct temporal stages of the experimental wave, variables not applicable to the specific record type are stored as nulls (represented here as dashes). The excerpt captures the 1--5 Likert responses for the psychometric items (\texttt{psyscale01Q1}--\texttt{Q24}), quantitative persuasion metrics (\texttt{feedbackQ0}--\texttt{Q3}), and the original qualitative text (\texttt{feedbackQ4}). Universally unique identifiers (UUIDs) ensure privacy-preserving data linkage.}
\label{tab:psyscales_sample}
\begin{tabular}{@{} l >{\raggedright\arraybackslash}p{4cm} >{\raggedright\arraybackslash}p{9cm} @{}}
\toprule
\textbf{Variable} & \textbf{Example 1} & \textbf{Example 2} \\ \midrule
userId & \path{a1983b67-b48c-4c21-a908-7abb84ed4ddd} & \path{8c154903-5aee-4361-b6f4-06496d7decdf} \\ \addlinespace
name & psyscales & feedback \\ \addlinespace
psyscaleId & \path{8406391e-9db7-475d-9de7-1558a90766f3} & \path{d2a41dad-3a95-4df2-b5b5-475a4d2d1db8} \\ \addlinespace
sessionId & \path{552e0988-b168-4724-8bd9-029cc1794792} & \path{b73decea-a0ef-4f16-808f-d1ffb446786d} \\ \addlinespace
psyscale01Q1 & 4 & - \\
psyscale01Q2 & 5 & - \\
\dots & \dots & \dots \\
psyscale01Q23 & 3 & - \\
psyscale01Q24 & 3 & - \\ \addlinespace
feedbackQ0 & - & 100 \\
feedbackQ1 & - & 50 \\
feedbackQ2 & - & 25 \\
feedbackQ3 & - & 93 \\
feedbackQ4 & - & Il cambiamento climatico si riferisce a variazioni a lungo termine delle temperature e dei modelli meteorologici. Queste variazioni possono essere naturali, ad esempio a causa di cambiamenti nell'attività solare o di grandi eruzioni vulcaniche. Tuttavia, a partire dal 1800, le attività umane sono diventate il principale motore del cambiamento climatico, principalmente a causa della combustione di combustibili fossili come carbone, petrolio e gas. Cause del cambiamento climatico: * Combustione di combustibili fossili: La combustione di carbone, petrolio e gas per produrre energia elettrica, calore e trasporti rilascia grandi quantità di anidride carbonica (CO2) e altri gas serra nell'atmosfera. Questi gas intrappolano il calore del sole, portando a un aumento delle temperature globali. * Deforestazione: Le foreste assorbono CO2 dall'atmosfera. Il taglio degli alberi non solo elimina questo importante pozzo di carbonio, ma il legno in decomposizione o bruciato rilascia anche il carbonio immagazzinato nell'atmosfera. * Agr \\ \addlinespace
insertDate & 2025-05-24 08:00:10 & 2025-05-28 07:22:34 \\ \bottomrule
\end{tabular}
\end{table}

To facilitate statistical analysis, we provide an aggregated tabular file named \texttt{data\_table.csv}. This compiled dataset consists of 3,080 rows representing the four interaction waves for each participant. It consolidates sociodemographic variables with session-specific metadata (topic and model assignments). Importantly, this table includes the standardized factor scores derived from the psychometric analysis (See Section Technical validation \ref{sec:tech_validation}), the unidimensional \texttt{TILLMI} score and the two Need for Cognition subfactors (\texttt{NFC\_seek} and \texttt{NFC\_diligence}). These latent traits are mapped directly against the post-interaction feedback metrics ($Q0$--$Q4$) for each corresponding session, providing a ready-to-use format for longitudinal and predictive modeling without requiring JSON parsing.

Comprising the core textual corpus, \texttt{talk2ai\_conversations.json} holds 3,080 records representing the four synchronous chat waves completed by each user. Each record links a dialogue to specific session and user identifiers through a \texttt{messages} array. This array documents every turn in strict chronological order, specifying the speaker role (user or assistant), the precise creation timestamp, and the raw textual content.

\begin{table}[ht!]
\centering
\footnotesize
\caption{\textbf{Illustrative sample of conversational interaction records.} This table provides a representative excerpt of two message entries from the raw \texttt{talk2ai\_conversations.json} file, documenting the multi-turn dialogues between participants and their assigned large language models (Step 3). The data structure utilizes universally unique identifiers (UUIDs) for privacy-preserving linkage across users (\texttt{userId}), longitudinal phases (\texttt{sessionId}), and specific dialogue instances (\texttt{conversationId}). Each record represents a single utterance, defined by a unique \texttt{messageId}, the chronological \texttt{msg\_timestamp}, and the interlocutor's \texttt{role} (either \texttt{user} or \texttt{assistant}). The \texttt{content} variable captures the raw, unedited Italian text exchanged during the session. Example 1 illustrates a user's prompt regarding climate change, whereas Example 2 displays a language model's generated counter-argument concerning healthcare systems and misinformation.}
\label{tab:conversations_sample}
\begin{tabular}{@{} l >{\raggedright\arraybackslash}p{6.3cm} >{\raggedright\arraybackslash}p{6.3cm} @{}}
\toprule
\textbf{Variable} & \textbf{Example 1} & \textbf{Example 2} \\ \midrule
sessionId & \path{9d5f7607-d3dd-48e6-8eea-66d5037b9c7e} & \path{84012abf-6ff3-4399-b871-83601b3b49b5} \\ \addlinespace
userId & \path{c549d78a-4066-4a9e-b8ac-12dcc8e64359} & \path{aac8ea52-d791-412a-96ef-cb233e59e139} \\ \addlinespace
conversationId & \path{ae444a9e-6319-40d0-8b1f-04cf018e933d} & \path{e79e7e24-14d2-49b4-82ef-6d6e73def149} \\ \addlinespace
conv\_start\_date & 2025-05-26 16:52:43 & 2025-05-09 14:13:19 \\ \addlinespace
messageId & \path{2qvddqCtddUwPuRo} & \path{msg-YN4FDDOZjuQMRtlM8HfREZ7b} \\ \addlinespace
role & \texttt{user} & \texttt{assistant} \\ \addlinespace
content & La concentrazione di CO2 provoca l’innalzamento globale della temperatura che a sua volta rende sempre più frequenti fenomeni di inondazioni, siccità, dissesto idrogeologico, diffusione di malattie, crisi dei sistemi agricoli, crisi idrica e estinzione di specie animali e vegetali. Non possiamo più attendere, dobbiamo invertire la rotta & Hai sollevato punti validi. È vero che le "finte evidenze" confondono, ma il problema non è solo il SSN: è sistemico. I social hanno un impatto enorme (vedi teorie complottiste virali), mentre l’istruzione, indebolita da scelte politiche, non prepara a discernere le fonti. Sul collasso del SSN, alcune regioni (es. Emilia-Romagna) hanno performance migliori, ma il finanziamento inadeguato e la burocrazia soffocano l’efficienza. Il privato è un tampone, ma il vero problema è la mancanza di investimenti strutturali. Gli anziani sono solo un sintomo: il sistema non incentiva l’aggiornamento continuo, nemmeno per i giovani. (100 parole) \\ \addlinespace
msg\_timestamp & 2025-05-26 16:54:54 & 2025-05-09 14:16:16 \\ \bottomrule
\end{tabular}
\end{table}

To enable comparative linguistic analysis, the \texttt{translations} folder provides these same 3,080 dialogues translated into English, Spanish, Dutch, German, Portuguese, and French. Stored as separate compressed archives for each language, these files maintain a one-to-one mapping with the original Italian identifiers, ensuring that the message sequence and role-play dynamics remain structurally identical across all seven languages.

\section*{Technical Validation} \label{sec:tech_validation}

To evaluate the structural validity of the psychometric instruments at baseline (Time 1), the sample was split into independent exploratory and confirmatory subsets using demographic-stratified sampling. Sample sizes varied across specific scale analyses due to construct-specific exclusion criteria. Specifically, 69 participants who responded negatively to the baseline \texttt{knowLlms} screening question (\textit{"Have you ever heard of language models like ChatGPT?"}) were excluded from the baseline validation of the TILLMI scale, as they could not validly report a pre-existing attitude toward an unknown technology. Consequently, the TILLMI models were validated on the remaining 701 participants (EFA subset $N=336$; CFA subset $N=365$). The Need for Cognition (NFC) scale, which measures a general cognitive trait, did not require this domain-specific filtering and was validated on the full baseline cohort (EFA subset $N=368$; CFA subset $N=402$). To preserve theoretical continuity, item selection for all analyses heavily favored the previously validated structures of these scales, tolerating minor statistical deviations where conceptually appropriate.

\subsection*{Exploratory Factor Analysis (EFA)}
For the TILLMI scale ($N=336$), initial diagnostics of the 8 items showed excellent overall suitability (Bartlett's $\chi^2(28) = 1014.96$, $p < .001$; KMO $MSA = .88$). However, Item 7 was removed due to inadequate sampling adequacy ($MSA = .54$) and problematic cross-loading. A subsequent EFA on the remaining 7 items confirmed high suitability (KMO $MSA = .89$). Although initial eigenvalues suggested a secondary factor, scree plot inspection and significant cross-loadings in forced two-factor solutions supported retaining a parsimonious, unidimensional 7-item model.

For the NFC scale ($N=368$), the 6 items demonstrated adequate suitability (Bartlett's $\chi^2(15) = 551.81$, $p < .001$; KMO $MSA = .75$). Item 3 exhibited marginal adequacy ($MSA = .52$) but was retained to preserve the established 6-item theoretical structure. Parallel and scree plot analyses indicated a two-factor solution explaining 53.9\% of the variance. Using an oblimin rotation, the items separated cleanly into Factor 1 (\texttt{NFC\_seek}: items 1, 2, 5, 6; loadings $> .68$) and Factor 2 (\texttt{NFC\_diligence}: items 3, 4; loadings $> .65$). This two-factor structure was retained as theoretically superior to a unidimensional model.

\subsection*{Confirmatory Factor Analysis (CFA)}
The EFA-derived models were tested on a separate confirmatory subset at Time 1 using the DWLS estimator for ordered categorical data.

The 7-item unidimensional TILLMI model showed excellent fit ($\CFI = .998$, $\TLI = .998$, $\SRMR = .034$, $\RMSEA = .043$ [90\% CI: $.000, .072$]). Internal consistency, calculated via ordinal Cronbach's alpha from the polychoric correlation matrix, was high ($\alpha = .902$, 95\% CI $[.73, .98]$).

The two-factor NFC model yielded robust global fit indices ($\CFI = .995$, $\TLI = .990$, $\SRMR = .041$, $\RMSEA = .067$ [90\% CI: $.035, .101$]), outperforming a competing 1-factor model ($\CFI = .834$). The model produced a warning of negative estimated residual variance for item 4, a known artifact for factors indicated by only two items. Ordinal Cronbach's alpha for \texttt{NFC-Seek} was strong ($\alpha = .855$). The \texttt{NFC-Diligence} factor fell marginally below the standard .70 threshold ($\alpha = .683$), which is statistically expected for a two-item construct and remains theoretically valuable.

\subsection*{Longitudinal Measurement Invariance}
Finally, the models were tested for longitudinal measurement invariance across all four waves using the full sample ($N=770$; Table \ref{tab:LMI_tillmi_nfc}). 

For both scales, the change in CFI from the metric to the scalar model remained within the accepted tolerance ($\Delta\CFI \le .011$). The TILLMI model achieved scalar invariance ($\Delta\CFI = -.010$) despite an elevated baseline configural $\RMSEA$ (.110), likely reflecting RMSEA's known sensitivity to model complexity. The NFC model showed acceptable configural fit ($\RMSEA = .085$) and passed scalar invariance ($\Delta\CFI = -.011$). Establishing scalar invariance confirms that the factor structures, loadings, and item intercepts remained equivalent across waves, ensuring the validity of latent score comparisons over the course of the longitudinal study.

\begin{table}[ht!]
\centering
\footnotesize
\caption{\textbf{Longitudinal measurement invariance fit indices for psychometric scales.} This table presents the model fit statistics establishing the structural consistency of the Trust in Large Language Models Instrument (TILLMI; 7-item, 1-factor) and the Need for Cognition scale (NFC; 6-item, 2-factor) across the four experimental waves (full sample, $N=770$). Configural, metric, and scalar invariance models were sequentially estimated to confirm that the psychometric properties of the instruments remained stable over repeated longitudinal assessments. All reported fit indices are robust scaled values (\texttt{.scaled}) extracted via the \texttt{lavaan} package. The $\Delta$CFI column indicates the change in CFI relative to the preceding, less constrained model (e.g., Metric vs. Configural). Abbreviations: CFI, Comparative Fit Index; TLI, Tucker-Lewis Index; SRMR, Standardized Root Mean Square Residual; RMSEA, Root Mean Square Error of Approximation; CI, Confidence Interval.}
\label{tab:LMI_tillmi_nfc}
\begin{tabular}{@{} >{\raggedright\arraybackslash}p{3.5cm} l c c c c l @{}}
\toprule
\textbf{Scale} & \textbf{Model} & \textbf{CFI} & \textbf{$\Delta$CFI} & \textbf{TLI} & \textbf{SRMR} & \textbf{RMSEA [90\% CI]} \\
\midrule
TILLMI & Configural & .891 & -- & .880 & .062 & .110 [.107, .114] \\
(7-item, 1-factor) & Metric & .918 & +.027 & .914 & .064 & .093 [.090, .096] \\
 & Scalar & .908 & -.010 & .922 & .064 & .089 [.086, .092] \\
\addlinespace
Need for Cognition & Configural & .945 & -- & .933 & .053 & .085 [.081, .089] \\
(6-item, 2-factor) & Metric & .955 & +.010 & .947 & .056 & .075 [.071, .080] \\
 & Scalar & .944 & -.011 & .950 & .056 & .073 [.070, .077] \\
\bottomrule
\end{tabular}
\end{table}

\section*{Usage Notes}\label{sec:data_usage}

To utilize the \textit{Talk2AI} dataset, researchers integrate the files using unique identifiers. The user demographic data (\path{talk2ai_registries.json}), participant metadata (\path{data_table.csv}), psychometric evaluations (\path{talk2ai_psyscales.json}), and conversational transcripts (\path{talk2ai_conversations.json}) merge at the user level using the \texttt{userId} key. For temporal analysis, researchers link records across \path{talk2ai_conversations.json} and \path{talk2ai_psyscales.json} using the \texttt{sessionId} key. Session records sort chronologically using session date values. This linkage permits the analysis of user responses and interactions across the experiment waves.

When analyzing user traits, researchers utilize standardized factor scores rather than computations of raw values or unweighted linear combinations of the Likert items. These scores originate from Longitudinal Measurement Invariance (LMI) testing to ensure measurement stability across waves (see Technical Validation). The Trust in Large Language Models Instrument (TILLMI) provides a unidimensional factor score derived from exploratory and confirmatory factor analyses (see Section Technical validation), supporting a single-factor solution for this dataset. However, the validation study \cite{deduro2025tillmi} posited a two-factor structure. The Need for Cognition (NFC) scale provides two subfactor scores: \texttt{NFC\_seek} (motivation for cognitive tasks) and \texttt{NFC\_diligence} (persistence in mental discipline, derived from reverse-coded items \texttt{psyscale01Q11} and \texttt{psyscale01Q12}). For both scales, in all factors, increases in standardized scores correspond to increases in the target trait.

By linking categorical variables (demographics, assigned topics, LLM architecture) and psychometric baselines with session feedback (\texttt{feedbackQ0}--\texttt{feedbackQ3}) and textual data (\texttt{content}, \texttt{feedbackQ4}), the dataset architecture facilitates multidimensional analytics. Researchers can apply Natural Language Processing (NLP) to the textual logs and feedback responses for feature engineering. Because the system prompts instructed the models to identify argument flaws, the resulting conversational data provides a foundation for fallacy detection, stance classification, and adversarial argumentation modeling. The native \texttt{content} variable consists of unedited Italian text, which requires Italian-compatible libraries (e.g., \texttt{spaCy} Italian models) for syntax parsing and extraction. Alternatively, the machine-translated English subsets accommodate English-centric computational tools. Finally, extracted features---such as syntax complexity, sentiment valence, and lexical diversity---can be cross-referenced with metadata to profile interaction styles, allowing researchers to test associations like the relationship between trust baselines and semantic density across different LLM architectures.

Text embeddings combined with factor scores support unsupervised learning methodologies. Concatenating numerical arrays with vector representations of text permits dimensionality reduction and clustering algorithms (e.g., K-Means, HDBSCAN). This approach clusters user typologies, isolating profiles that display linguistic compliance alongside non-persuasion. Unsupervised topic modeling applied across waves tracks semantic drift and argument evolution within the experiment topics.

Baseline traits and textual sequences provide the feature space for supervised predictive modeling. Researchers frame persuasion outcomes as regression or classification tasks using demographics, factor scores, and NLP features as independent variables. These feature sets train algorithms to forecast outcomes. Target variables include classification of stance shifts or regression of persuasion metrics and model humanness evaluations (\texttt{feedbackQ2}).

The longitudinal design supports sequence modeling. Transforming conversational data into time-series sequences anchored with session feedback allows researchers to train Recurrent Neural Networks (RNNs) or Markov chain models to predict argumentative trajectories. These models calculate how dialogue strategies associate with the stability of user convictions. The psychometric variables permit Structural Equation Modeling (SEM) to test whether NLP features mediate the relationship between baseline traits and persuasion outcomes.

\section*{Data Availability} \label{sec:data_availability}

The Talk2AI dataset is publicly available in the GitHub repository at \url{https://github.com/MassimoStel/Talk2AI/tree/main/Data_paper}. The repository provides a primary \texttt{Data\_files.zip} archive containing the JSON datasets (\texttt{talk2ai\_registries.json}, \texttt{talk2ai\_psyscales.json}, and \texttt{talk2ai\_conversations.json}). Additionally, a \texttt{translations} directory is provided, containing the conversation logs computationally translated from the original Italian into English, Spanish, Dutch, German, Portuguese, and French.

\section*{Code Availability}\label{sec:code_availability}

The computational logic used to derive the final cohort is documented in the \path{talk2ai_1_1_data_extraction.ipynb} Jupyter notebook, available within the GitHub repository (\url{https://github.com/MassimoStel/Talk2AI/tree/main/Data_paper}). This script outlines the steps taken in Python to filter the initial sample of 2,644 registered participants down to the 770 users who met the quality control and longitudinal completion criteria. Exploratory factor analysis, and confirmatory factor analysis for technical validation were conducted using the \texttt{lavaan} and \texttt{psych} packages in R, as presented in the \path{talk2ai_2_2_psychometric_FA.ipynb} file. Longitudinal measurement invariance and structural equation modeling can be accessed in \path{talk2ai_2_3_longitudinal_analysis.ipynb}.

\bibliographystyle{naturemag}
\bibliography{references}

@article{brady2025dual,
  title={Dual-process theory and decision-making in large language models},
  author={Brady, Oliver and Nulty, Paul and Zhang, Lili and Ward, Tom{\'a}s E and McGovern, David P},
  journal={Nature Reviews Psychology},
  pages={1--16},
  year={2025},
  publisher={Nature Publishing Group US New York}
}

@article{argyle2023leveraging,
  title={Leveraging AI for democratic discourse: Chat interventions can improve online political conversations at scale},
  author={Argyle, Lisa P and Bail, Christopher A and Busby, Ethan C and Gubler, Joshua R and Howe, Thomas and Rytting, Christopher and Sorensen, Taylor and Wingate, David},
  journal={Proceedings of the National Academy of Sciences},
  volume={120},
  number={41},
  pages={e2311627120},
  year={2023},
  publisher={National Academy of Sciences}
}

@article{matz2024potential,
  title={The potential of generative AI for personalized persuasion at scale},
  author={Matz, Sandra C and Teeny, Jacob D and Vaid, Sumer S and Peters, Heinrich and Harari, Gabriella M and Cerf, Moran},
  journal={Scientific Reports},
  volume={14},
  number={1},
  pages={4692},
  year={2024},
  publisher={Nature Publishing Group UK London}
}

@article{carrasco2024large,
  title={Large language models are as persuasive as humans, but how? About the cognitive effort and moral-emotional language of LLM arguments},
  author={Carrasco-Farre, Carlos},
  journal={arXiv preprint arXiv:2404.09329},
  year={2024}
}

@article{bai2025llm,
  title={LLM-generated messages can persuade humans on policy issues},
  author={Bai, Hui and Voelkel, Jan G and Muldowney, Shane and Eichstaedt, Johannes C and Willer, Robb},
  journal={Nature Communications},
  volume={16},
  number={1},
  pages={6037},
  year={2025},
  publisher={Nature Publishing Group UK London}
}

@article{goldstein2024persuasive,
  title={How persuasive is AI-generated propaganda?},
  author={Goldstein, Josh A and Chao, Jason and Grossman, Shelby and Stamos, Alex and Tomz, Michael},
  journal={PNAS nexus},
  volume={3},
  number={2},
  pages={pgae034},
  year={2024},
  publisher={Oxford University Press US}
}

@article{simchon2024persuasive,
  title={The persuasive effects of political microtargeting in the age of generative artificial intelligence},
  author={Simchon, Almog and Edwards, Matthew and Lewandowsky, Stephan},
  journal={PNAS nexus},
  volume={3},
  number={2},
  pages={pgae035},
  year={2024},
  publisher={Oxford University Press US}
}

@article{hackenburg2025levers,
  title={The levers of political persuasion with conversational AI},
  author={Hackenburg, Kobi and Tappin, Ben M and Hewitt, Luke and Saunders, Ed and Black, Sid and Lin, Hause and Fist, Catherine and Margetts, Helen and Rand, David G and Summerfield, Christopher},
  journal={arXiv preprint arXiv:2507.13919},
  year={2025}
}

@article{salvi2025conversational,
  title={On the conversational persuasiveness of GPT-4},
  author={Salvi, Francesco and Horta Ribeiro, Manoel and Gallotti, Riccardo and West, Robert},
  journal={Nature Human Behaviour},
  pages={1--9},
  year={2025},
  publisher={Nature Publishing Group UK London}
}

@article{argyle2025testing,
  title={Testing theories of political persuasion using AI},
  author={Argyle, Lisa P and Busby, Ethan C and Gubler, Joshua R and Lyman, Alex and Olcott, Justin and Pond, Jackson and Wingate, David},
  journal={Proceedings of the National Academy of Sciences},
  volume={122},
  number={18},
  pages={e2412815122},
  year={2025},
  publisher={National Academy of Sciences}
}

@article{rossetti2024social,
  title={Y social: an llm-powered social media digital twin},
  author={Rossetti, Giulio and Stella, Massimo and Cazabet, R{\'e}my and Abramski, Katherine and Cau, Erica and Citraro, Salvatore and Failla, Andrea and Improta, Riccardo and Morini, Virginia and Pansanella, Valentina},
  journal={arXiv preprint arXiv:2408.00818},
  year={2024}
}

@misc{deduro2025tillmi,
      title={Measuring and identifying factors of individuals' trust in Large Language Models}, 
      author={Edoardo Sebastiano De Duro and Giuseppe Alessandro Veltri and Hudson Golino and Massimo Stella},
      year={2025},
      eprint={2502.21028},
      archivePrefix={arXiv},
      primaryClass={cs.HC},
      url={https://arxiv.org/abs/2502.21028}, 
}

@article{watson2024ai,
  author  = {Watson, Jared and Valsesia, Francesca and Segal, Shoshana},
  title   = {Assessing {AI} receptivity through a persuasion knowledge lens},
  journal = {Current Opinion in Psychology},
  year    = {2024},
  volume  = {58},
  pages   = {101834},
  doi     = {10.1016/j.copsyc.2024.101834}
}

@article{huang2023persuasive,
  author  = {Huang, Guanxiong and Wang, Sai},
  title   = {Is artificial intelligence more persuasive than humans? A meta-analysis},
  journal = {Journal of Communication},
  year    = {2023},
  volume  = {73},
  number  = {6},
  pages   = {552--562},
  doi     = {10.1093/joc/jqad024}
}

@article{blankenship2023certainty,
  author  = {Blankenship, Kevin L. and Machacek, Marielle G. and Standefer, Jack},
  title   = {Resistance strategies and attitude certainty in persuasion: bolstering vs. counterarguing},
  journal = {Frontiers in Psychology},
  year    = {2023},
  volume  = {14},
  pages   = {1191293},
  doi     = {10.3389/fpsyg.2023.1191293}
}

@misc{openai2024gpt4ocard,
      title={GPT-4o System Card}, 
      author={OpenAI and : and Aaron Hurst and Adam Lerer and Adam P. Goucher and Adam Perelman and Aditya Ramesh and others},
      year={2024},
      eprint={2410.21276},
      archivePrefix={arXiv},
      primaryClass={cs.CL},
      url={https://arxiv.org/abs/2410.21276}, 
}

@techreport{claudesonnet37systemcard,
  author      = {{Anthropic}},
  title       = {Claude 3.7 Sonnet System Card},
  institution = {Anthropic},
  year        = {2025},
  month       = {February},
  url         = {https://www-cdn.anthropic.com/9ff93dfa8f445c932415d335c88852ef47f1201e.pdf},
  note        = {Accessed: 2025-02-24}
}

@misc{mistral2024large,
  author       = {{Mistral AI}},
  title        = {Mistral Large: Our New Flagship Model},
  howpublished = {Mistral AI Blog},
  year         = {2024},
  month        = {February},
  url          = {https://mistral.ai/news/mistral-large},
  note         = {Accessed: 2025-02-24}
}

@Article{Adler2000MacArthurScale,
    author={Adler, Nancy E.
    and Epel, Elissa S.
    and Castellazzo, Grace
    and Ickovics, Jeannette R.},
    title={Relationship of subjective and objective social status with psychological and physiological functioning: Preliminary data in healthy, White women.},
    journal={Health Psychology},
    year={2000},
    publisher={American Psychological Association},
    address={US},
    volume={19},
    number={6},
    pages={586-592},
    doi={10.1037/0278-6133.19.6.586},
    url={https://doi.org/10.1037/0278-6133.19.6.586}
}

@article{deHolanda2020NFC6items,
    author = {Gabriel Lins de Holanda Coelho and Paul H. P. Hanel and Lukas J. Wolf},
    title ={The Very Efficient Assessment of Need for Cognition: Developing a Six-Item Version*},
    journal = {Assessment},
    volume = {27},
    number = {8},
    pages = {1870-1885},
    year = {2020},
    doi = {10.1177/1073191118793208},
    note ={PMID: 30095000},
    URL = {https://doi.org/10.1177/1073191118793208},
    eprint = {https://doi.org/10.1177/1073191118793208}
}

@article{Rammstedt2007BFQ10,
    title = {Measuring personality in one minute or less: A 10-item short version of the Big Five Inventory in English and German},
    journal = {Journal of Research in Personality},
    volume = {41},
    number = {1},
    pages = {203-212},
    year = {2007},
    issn = {0092-6566},
    doi = {https://doi.org/10.1016/j.jrp.2006.02.001},
    url = {https://www.sciencedirect.com/science/article/pii/S0092656606000195},
    author = {Beatrice Rammstedt and Oliver P. John}
}

@Article{Cartwright2023DictatorDonation,
    author={Cartwright, Edward
    and Thompson, Adam},
    title={Using Dictator Game Experiments to Learn About Charitable Giving},
    journal={VOLUNTAS: International Journal of Voluntary and Nonprofit Organizations},
    year={2023},
    month={Feb},
    day={01},
    volume={34},
    number={1},
    pages={185-191},
    issn={1573-7888},
    doi={10.1007/s11266-022-00490-7},
    url={https://doi.org/10.1007/s11266-022-00490-7}
}

@article{liu2024deepseek,
  title={Deepseek-v3 technical report},
  author={Liu, Aixin and Feng, Bei and Xue, Bing and Wang, Bingxuan and Wu, Bochao and Lu, Chengda and Zhao, Chenggang and Deng, Chengqi and Zhang, Chenyu and Ruan, Chong and others},
  journal={arXiv preprint arXiv:2412.19437},
  year={2024}
}

\section*{Acknowledgements}

The authors acknowledge Salvatore Citraro for preparing the CSV files for the data translations.

\section*{Author Contributions}

M.S. and G.A.V. conceptualized the study, developed the theoretical framework, designed the methodology, and acquired the funding. M.S. acted as the principal investigator, providing resources and overall supervision with support from G.A.V. E.T. developed the custom experimental software platform and executed the data gathering and data curation processes. A.C. and M.S. conducted the formal data and psychometric analyses. A.C., A.A.A. and M.S. drafted the manuscript. All authors reviewed, edited, and approved the final manuscript.

\section*{Competing Interests}
The authors declare no competing interests.

\section*{Funding}

The authors acknowledge support from the following grants: Call for Research Grant 2023 funded by University of Trento (ID: PS 22\_27, A.C., E.T., G.A.V. and M.S.); CALCOLO project funded by Fondazione VRT (M.S.); FIS project funded by Ministero dell'Università e della Ricerca, D.D.N. 23178, 10/12/2024, BANDO FIS2, ID: FIS-2023-02086 (M.S.).

\end{document}